\documentclass[twocolumn]{aastex631}
\usepackage[version=4]{mhchem}
\usepackage{xspace}
\usepackage{hyperref}
\usepackage{cleveref}

\accepted{29 August 2023}
\submitjournal{ApJ}

\graphicspath{{./}{Figures/}}

\begin{document}

\newcommand{\Rp}{$R_\mathrm{p}$\xspace}
\newcommand{\Tp}{$T_\mathrm{p}$\xspace}
\newcommand{\Rs}{$R_\mathrm{s}$\xspace}
\newcommand{\Ts}{$T_\mathrm{s}$\xspace}

\newcommand{\R}[1]{$R_\mathrm{#1}$\xspace}
\newcommand{\T}[1]{$T_\mathrm{#1}$\xspace}

\title{On the Effectiveness of the Planetary Infrared Excess (PIE) Technique to Retrieve the Parameters of Multiplanet Systems around M dwarfs: A Case Study on the TRAPPIST-1 System.}

\shorttitle{TRAPPIST-1 PIE}
\shortauthors{Mayorga and CHAMPs Team}

\correspondingauthor{L. C. Mayorga}
\email{laura.mayorga@jhuapl.edu}

\author[0000-0002-4321-4581]{L. C. Mayorga}
\affiliation{The Johns Hopkins University Applied Physics Laboratory, 11100 Johns Hopkins Rd, Laurel, MD, 20723, USA}
\author[0000-0002-0746-1980]{J. Lustig-Yaeger}
\affiliation{The Johns Hopkins University Applied Physics Laboratory, 11100 Johns Hopkins Rd, Laurel, MD, 20723, USA}
\affiliation{NASA NExSS Virtual Planetary Laboratory, Box 351580, University of Washington, Seattle, Washington 98195, USA}
\author[0000-0002-7352-7941]{K. B. Stevenson}
\affiliation{The Johns Hopkins University Applied Physics Laboratory, 11100 Johns Hopkins Rd, Laurel, MD, 20723, USA}
\collaboration{3}{Consortium on Habitability and Atmospheres of M-dwarf Planets (CHAMPs)}

\begin{abstract}
The planetary infrared excess (PIE) technique has the potential to efficiently detect and characterize the thermal spectra of both transiting and non-transiting exoplanets. However, the technique has not been evaluated on multiplanet systems. We use the TRAPPIST-1 system as our test bed to evaluate PIE's ability to resolve multiple planets. We follow the unfolding discoveries in the TRAPPIST-1 system and examine the results from the PIE technique at every stage. We test the information gained from observations with JWST and next-generation infrared observatories like the proposed MIRECLE mission concept. We find that even in the case where only the star is known, the PIE technique would infer the presence of multiple planets in the system. The precise number inferred is dependent on the wavelength range of the observation and the noise level of the data. We also find that in such a tightly packed, multiplanet system such as TRAPPIST-1, the PIE technique struggles to constrain the semi-major axis beyond prior knowledge. Despite these drawbacks and the fact that JWST is less sensitive to the fluxes from planets g and h, with strong priors in their orbital parameters we are able to constrain their equilibrium temperatures. We conclude that the PIE technique may enable the discovery of unknown exoplanets around solar-neighborhood M dwarfs and could characterize known planets around them.
\end{abstract}

\keywords{Exoplanets(498); Exoplanet detection methods (489); Extrasolar rocky planets(511); Infrared excess(788)}

\section{Introduction} \label{sec:intro}
There remain many hurdles in the search for habitable planets around other stars. While terrestrial planets around Sun-like stars are more akin to the inhabited planet in our Solar System, the signal size caused by the atmospheric features of a terrestrial planet around an M dwarf are much more favorable for detection \citep{Tarter2007AStars}. Thus M dwarfs as host stars have drawn additional study and critique \citep{Segura2010TheDwarf, Shields2016, Ranjan2017TheFollow-up, Claudi2021Super-earthsLab}. M dwarfs are also a compelling search target because they comprise nearly 75\% of the stars in the Solar neighborhood. While the number of terrestrial exoplanets discovered by space- and ground-based surveys now numbers in the thousands, the comparative number of habitable zone exoplanets and candidates within 30\,pc of Earth numbers only five \citetext{including the TRAPPIST-1 planets, and all around M dwarfs, NASA Exoplanet Archive}.

This known population is similar in number to that predicted by theory \citep[seven planets within 15\,pc;][]{Barclay2018}. Thus the search for life is stymied by the small sample size of the known population. Our reliance on techniques such as transmission and emission spectroscopy/photometry has constrained the population to only transiting planets. However, if we remove the transiting requirement, \citet{Barclay2018} predict nearly 240 non-transiting habitable zone terrestrial planets within 15\,pc. 

\begin{figure}
    \centering
    \includegraphics[width=\linewidth]{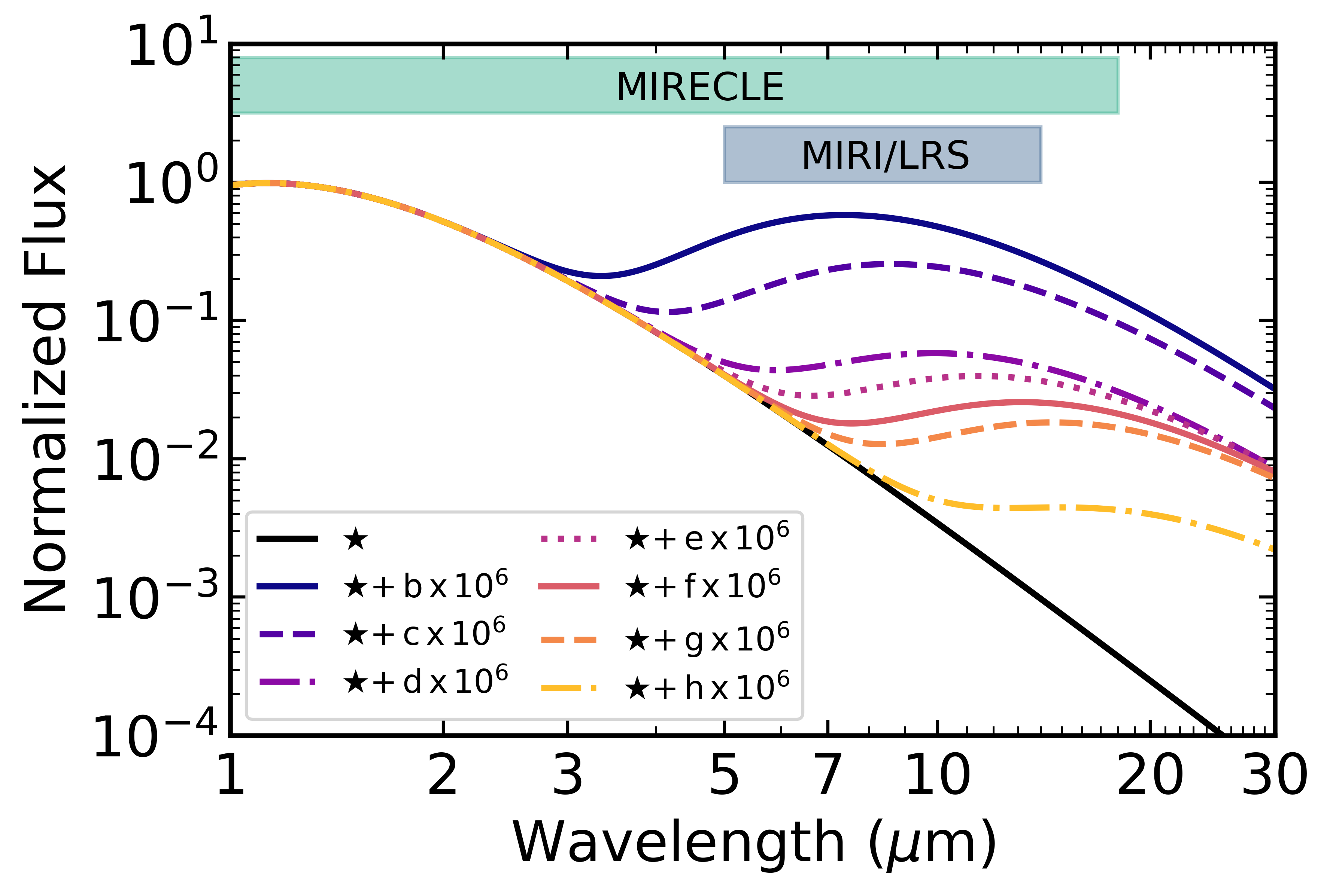}
    \caption{The PIE technique exemplified on the TRAPPIST-1 system assuming all objects in the system are blackbodies. We scale up the planet flux by $10^6$ for demonstration purposes and overlay the wavelength coverage of present and proposed IR missions.}
    \label{fig:toy}
\end{figure}

Building on the ideas of detecting circumstellar disks around stars, \citet{Stevenson2020} presented the concept of using a planet's infrared signal to determine its radius and temperature, particularly powerful in non-transiting planet scenarios. The technique takes advantage of the fact that the planet's peak flux is separated in wavelength space from the star's peak flux, making it detectable and characterizable even in non-transiting scenarios. We demonstrate this technique in \autoref{fig:toy} for the TRAPPIST-1 system. The initial demonstrations with a simulated WASP-43 and Proxima Centauri showed the technique had promise under the blackbody assumption for recovering a planet's radius and temperature. Concerns over any radius-temperature degeneracies were found to be mitigated by broad wavelength coverage. \citet{Stevenson2020} also identified a number of obstacles to the Planetary Infrared Excess (PIE) technique, including the concern of degeneracies presented by attempting to characterize planets in multiplanet systems. 
Since the initial formulation, the PIE technique has also been simulated for hot Jupiters in single-planet systems \citep{Lustig-Yaeger2021} and rocky exoplanets in white-dwarf systems \citep{Limbach2022}, both of which show promise for observations with JWST. \citet{Mandell2022} went further with simulated Earth-like planetary atmospheres in single-planet systems.
Using the TRAPPIST-1 system, we explore both transiting and non-transiting cases to determine the potential limitations of PIE while assuming blackbodies. 

To demonstrate the knowledge gained from the PIE technique, we present the performance using two observatory architectures, JWST and MIR Exoplanet CLimate Explorer \citep[MIRECLE;][a MIDEX mission concept featuring broad IR wavelength coverage]{Staguhn2019}. We follow the known history of TRAPPIST-1 as it unfolded and examine how well the PIE technique constrains the planetary parameters with the additional information gained over time. In \autoref{sec:method}, we outline how we built the TRAPPIST-1 dataset and the retrieval framework. Next, in \autoref{sec:results} we walk through TRAPPIST-1's discovery history and use the known information during each time period (\autoref{sec:star}, \autoref{sec:bcd}, \autoref{sec:spitzer}) as priors on our retrieval. Finally, we discuss the implications of our results in \autoref{sec:discuss} before concluding in \autoref{sec:conc}.

\section{Methods} \label{sec:method}
We follow the TRAPPIST-1 system discovery storyline, testing PIE at each stage of information gained: first with only knowledge of the star, TRAPPIST-1; then after the discovery of planets b, c, and a candidate ``d" \citep{Gillon2016}; then after the Spitzer observations with knowledge of all seven planets \citep{Gillon2017}.

We use the latest TRAPPIST-1 system parameters from \citet{Agol2021} as the true system parameters as shown in Table 1, but take the state of knowledge at the time to set priors. We assume Gaussian priors on the stellar parameters from the appropriate reference throughout this work and take the appropriate Gaussian or uniform prior for planetary parameters as they are known. For example, when planets b and c become known, we use Gaussian priors for their semi-major axes, $a$, and radii \Rp, but uniform priors on their Bond albedos, $A_B$. For any unconstrained planets, we use uniform priors for all properties. 

\begin{deluxetable}{l|CCCC}
\tablecaption{Table of Assumed ``True" System Parameters, Primarily Sourced from \citet{Agol2021}. \label{tbl:trap}}
\tablehead{\colhead{Body}\hspace{.25cm} & \colhead{\hspace{.75cm}$a$}\hspace{.75cm} & \colhead{\hspace{.5cm}$R$}\hspace{.5cm} & \colhead{\hspace{.5cm}$A_B$}\hspace{.5cm} & \colhead{\hspace{.5cm}$T$}\hspace{.5cm} \\
\cline{2-5} & \mathrm{(AU)} & (R_{\odot}, R_{\oplus}) & & \mathrm{(K)}}
\startdata
A & \nodata & 0.1192 & \nodata & 2566 \\
\hline
b & 0.01154 & 1.116 & 0.1 & 387 \\
c & 0.0158 & 1.097 & 0.1 & 331 \\
d & 0.02227 & 0.788 & 0.1 & 279 \\
e & 0.02925 & 0.92 & 0.1 & 243 \\
f & 0.03849 & 1.045 & 0.1 & 212 \\
g & 0.04683 & 1.129 & 0.1 & 192 \\
h & 0.06189 & 0.755 & 0.1 & 167 \\
\enddata
\tablecomments{We fix the Bond albedos and thus, calculate the equilibrium temperatures of the planets.}
\end{deluxetable}

For modeling, we limit our resolving power to $R$=100 and wavelength range to 1---18\,$\mu$m to mimic an expected MIRECLE \citep{Staguhn2019} observation (we will later explore a JWST configuration in \autoref{sec:jwst}). MIRECLE has already been shown to be effective for the PIE technique using realistic planetary atmospheres in single-planet systems \citep{Mandell2022}. We created the wavelength grid using the \texttt{create\_grid} function out of the PICASO package \citep{Batalha2019}. This wavelength range should capture the peak of the blackbody spectrum for objects in the range of about 2900--160\,K. We assume blackbody representations for all the relevant bodies. Thus, the system's flux is built up iteratively:
\begin{equation}
F = \sum_{i=A}^h \pi R_i^2 B_\lambda(T_i),
\end{equation}
where the equilibrium temperature of the planets, $T_{i=b-h}$, are computed from their semi-major axis, $a$, and Bond albedo, $A_B$, which we will hereafter refer to as just albedo.
\begin{equation}
T_{i=b-h} = T_A (1-A_B)^{1/4}\sqrt{\frac{R_A}{2a}},
\end{equation}
where $A$ subscripts denote stellar parameters. We choose to fit for $A_B$ and $a$ instead of $T$ partly to make use of their well-defined physical priors. We discuss this choice further in \autoref{sec:discuss}. To avoid identity issues while fitting, we do not allow planets to change their ordinal positions, i.e. $a_i < a_j$ $\forall$ $i$, $j$, where $i$ is the planet interior to $j$ ($a_b < a_c < a_d ... < a_N$). We do not evaluate the stability of the resulting orbits.

We use dynesty \citep{Speagle2020, dynesty} to infer the posterior probability distribution by fitting our blackbody model to the same simulated data using a $\chi^2$ likelihood function. The ``best fit" is taken to be the fit with the maximum likelihood or roughly the median posterior state vector. Dynesty, a nested sampling scheme, traverses nested shells of likelihood with a number of ``live points," which we set to roughly 100 times the number of fit parameters, to converge on high-likelihood regions of the prior volume. The ``data" is noiseless to avoid specific noise realizations from affecting our exploration of biases, but we impose an artificial noise floor\footnote{Over the course of this work, we discovered that under scenarios as precise and degenerate as these, dynesty struggled to converge to a solution and frequently raised an \texttt{AssertionError}. Subsequent trials showed that there was an underlying randomness to the failure rate that was partially alleviated for cases with increased the measurement uncertainties, modified priors (for cases with the solution near a parameter bound), reduced number of live points, or a different sampling method (we ultimately chose \texttt{rslice}). However, simply rerunning the failed simulation with a different random state of initialization was sufficient to achieve fully converged solutions.}, $\sigma_\mathrm{floor}$, in the likelihood function equivalent to 5\,ppm:
\begin{equation}
    \log L = -0.5 \sum_x{\left(\frac{Y_x-M_x}{\sigma_\mathrm{floor}}\right)^2},
\end{equation}
where $Y$ is the ``data" and $M$ is the model.

We evaluate best-fit models in two ways: (1) by examining the residuals in comparison to the expected noise and (2) by comparing the evidence, $\ln Z$. Unlike Markov Chain Monte Carlo algorithms, whose primary result is the posteriors, a nested sampling algorithm's first goal is to compute the evidence, and the posteriors are a useful by-product. Comparing the evidence of two models allows us to compute the Bayes factor ($\ln B = \Delta\ln Z$) and thus, rule out models (i.e., numbers of planets) by 3$\sigma$ \citep[when $\ln B > 3$;][]{Benneke2013HowSuper-earths}.

\section{Results} \label{sec:results}
\subsection{2015: Planet-free TRAPPIST-1A} \label{sec:star}
\begin{figure}
    \centering
    \includegraphics[width=\linewidth]{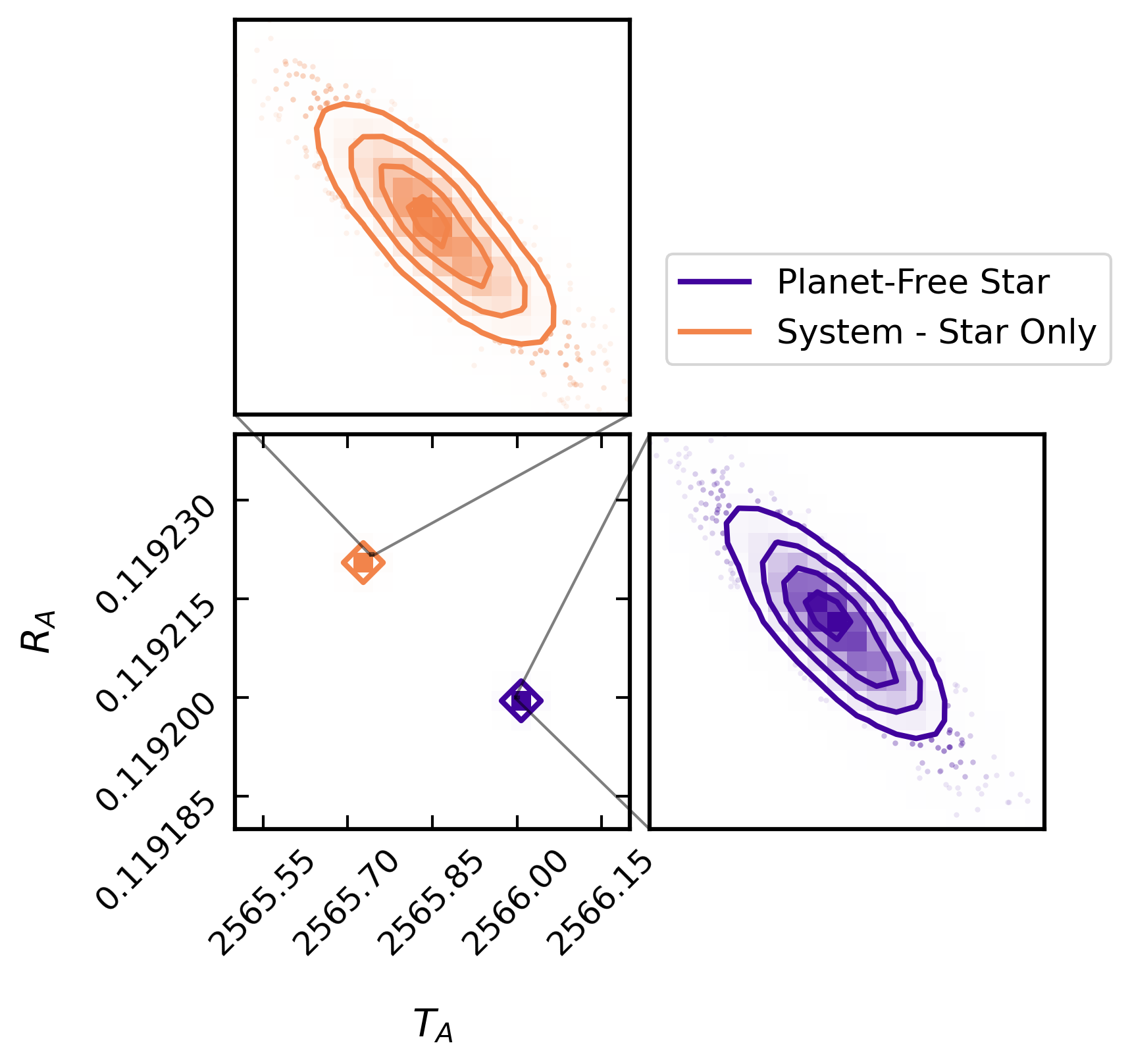}
    \includegraphics[width=\linewidth]{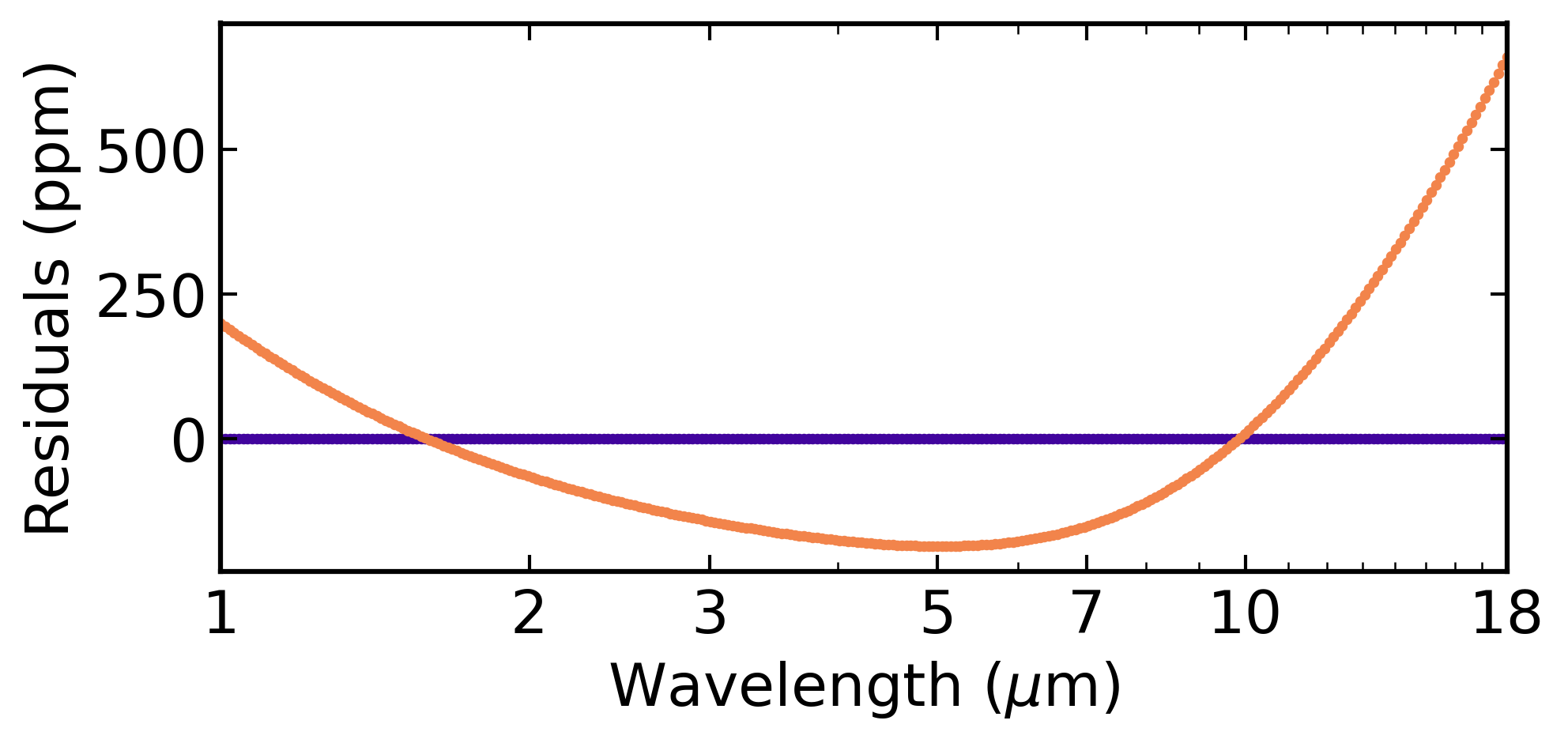}
    \caption{Posterior distribution and residuals for the star-only models retrieved with the best known parameters in 2015 from \citet{Filippazzo2015FUNDAMENTALREGIME} as Gaussian priors. Top: 2D posterior distributions for the stellar parameters. The zoomed insets show the correlation between temperature and radius more clearly. In the noiseless case here, the best-fit parameters are extremely precise. Bottom: the residuals of the best fit vs. the original model. In the ``Planet-Free" case, we fit a dataset of only TRAPPIST-1A. In ``System - Star Only," the dataset includes all planets. Thus, the PIE technique can distinguish between a solitary star and one hosting planets.}
    \label{fig:zero}
\end{figure}

We first take the parameters of the star as they were ``known" before the discovery of any planets in the TRAPPIST-1 system: 1.14$\pm$0.04\,\R{J} (0.117$\pm$0.004\,\R{\odot}) and 2557$\pm$64\,K. These would be the parameters provided by \citet{Filippazzo2015FUNDAMENTALREGIME} for what was then known as 2MASS\,J23062928-0502285 and are consistent with \citet{Agol2021} to 1$\sigma$. For demonstration, we start by taking the ``true" stellar parameters to simulate a planet-free TRAPPIST-1A from $1<\lambda<18\,\mu$m, with R=100 and $\sigma_\mathrm{floor}\equiv$\,5\,ppm. We test the performance of PIE with the constraints from each time period. We take our priors to be Gaussian centered on the known parameters from \citet{Filippazzo2015FUNDAMENTALREGIME} and show the posteriors in \autoref{fig:zero}. With our noiseless premise, a perfect fit would be represented by a $\chi^2$ of 0. The resulting $\chi^2$ is 0.00075 and the best-fit parameters match those of \citet{Agol2021} to high precision. The residuals are approximately 0\,ppm, i.e. well within our 5\,ppm noise floor, suggesting no additional blackbody components are necessary.

With the knowledge that we can precisely fit the case of a planet-free star, we now undertake the real TRAPPIST-1 system. We include all seven planets in the model using the ``true" parameters with the same noise floor, wavelength range, and spectral resolution. We again fit for just a star using the ``known" parameters, Gaussian posteriors. We show the results in \autoref{fig:zero}, including the normalized residuals in parts per million ((data-fit)/data$\times$10$^6$) as compared to our noise floor.
While the parameters of the star are well constrained, the residuals exceed 500 ppm at the longest wavelengths, and the $\chi^2$ value is extremely large.  Taken together, the evidence suggests the star is not the only source of flux. The best-fit parameters attempted to account for this flux by pushing the retrieved parameters to a colder (by 0.25\,K) and larger (by $2\times10^{-5}$\,\R{\odot}) star.

We are thus motivated to search for additional blackbody components in this dataset. 

\subsubsection{The Search for Planets}
We start with the addition of a planet dubbed planet ``b." We maintain Gaussian priors on the stellar parameters and now we add uniform priors for the planet's semi-major axis, radius, and albedo. We set the prior radius to search for planets between 0.5 and 4\,\R{\oplus} arbitrarily and prior albedo from 0 to 1.
\citet{Stevenson2020} emphasized the importance of fitting the peak of the blackbody flux to achieve the best constraints. This means we would be less sensitive to planets whose peaks fall outside of our wavelength range and use this to set the prior on the semi-major axis.
We compute the temperatures and the associated semi-major axes where capturing the peak of the planet is possible within our wavelength range. This leads to a uniform prior when blindly searching for planets at semi-major axes of 0.000212\,AU -- 0.068709\,AU (about 2900--160\,K).

\begin{figure}
    \centering
    \includegraphics[width=\linewidth]{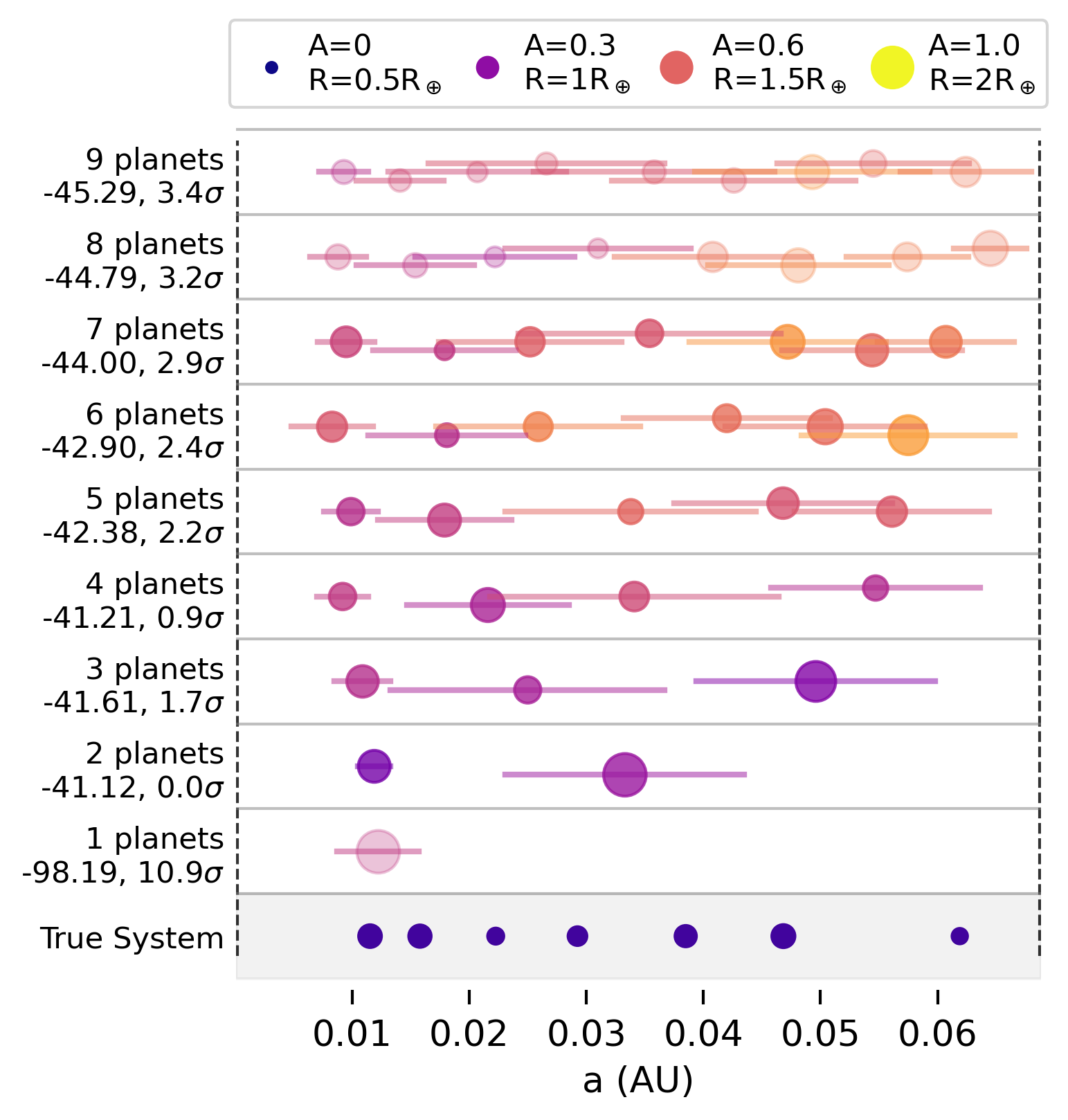}
    \caption{Best-fit system in each modeling scenario compared to the true TRAPPIST-1 system using only \citet{Filippazzo2015FUNDAMENTALREGIME} priors. The symbol size corresponds to the planet size, and the color indicates albedo. Horizontal bars indicate the uncertainty on a planet's semi-major axis. Each modeling scenario is labeled with the number of planets included, the computed evidence, and its significance against the best-fit model. We use transparency to demonstrate which models are ruled out by $>3\sigma$ confidence. The vertical dashed lines indicate the limits of our semi-major axis bounds.}
    \label{fig:2015mods}
\end{figure}

We add planets in the system until the evidence begins to rule out models. We take the model scenario with the highest evidence (in this case two planets) and compute the Bayes factor with all other scenarios (one to nine planets). The best fit of each scenario is shown in \autoref{fig:2015mods}. The favored model is two planets in the system (as supported by comparing evidences/Bayes Factor). Since the one-, eight-, and nine-planet cases are ruled out at more than 3$\sigma$ confidence, we can predict that there are two to seven planets in the TRAPPIST-1 system using the PIE technique without any prior information on the planets.

Ultimately since the best model is two planets, we can guide searches for planets around TRAPPIST-1 by indicating what periods we expect the planets to have. We use the best known mass of the star in 2015, 0.082\,$M_\odot$ \citep{Filippazzo2015FUNDAMENTALREGIME}, and assume circular orbits. The best-fit periods for a suspected planet ``b" is 1.65$^{+0.30}_{-0.28}$\,days with planet ``c" at 7.75$^{+3.8}_{-3.3}$\,days (the \citealp{Agol2021} values are 1.510826 and 2.421937, respectively). Despite the various planets included in each model, the stellar parameters are very robust. To correspond with our two-planet model, we would report stellar parameters of 0.1192\,$R_\odot$ and 2566\,K, which match \citet{Agol2021}.

\subsection{2016: Discovery of Planets b, c, and candidate ``d"} \label{sec:bcd}
\begin{figure}
    \centering
    \includegraphics[width=\linewidth]{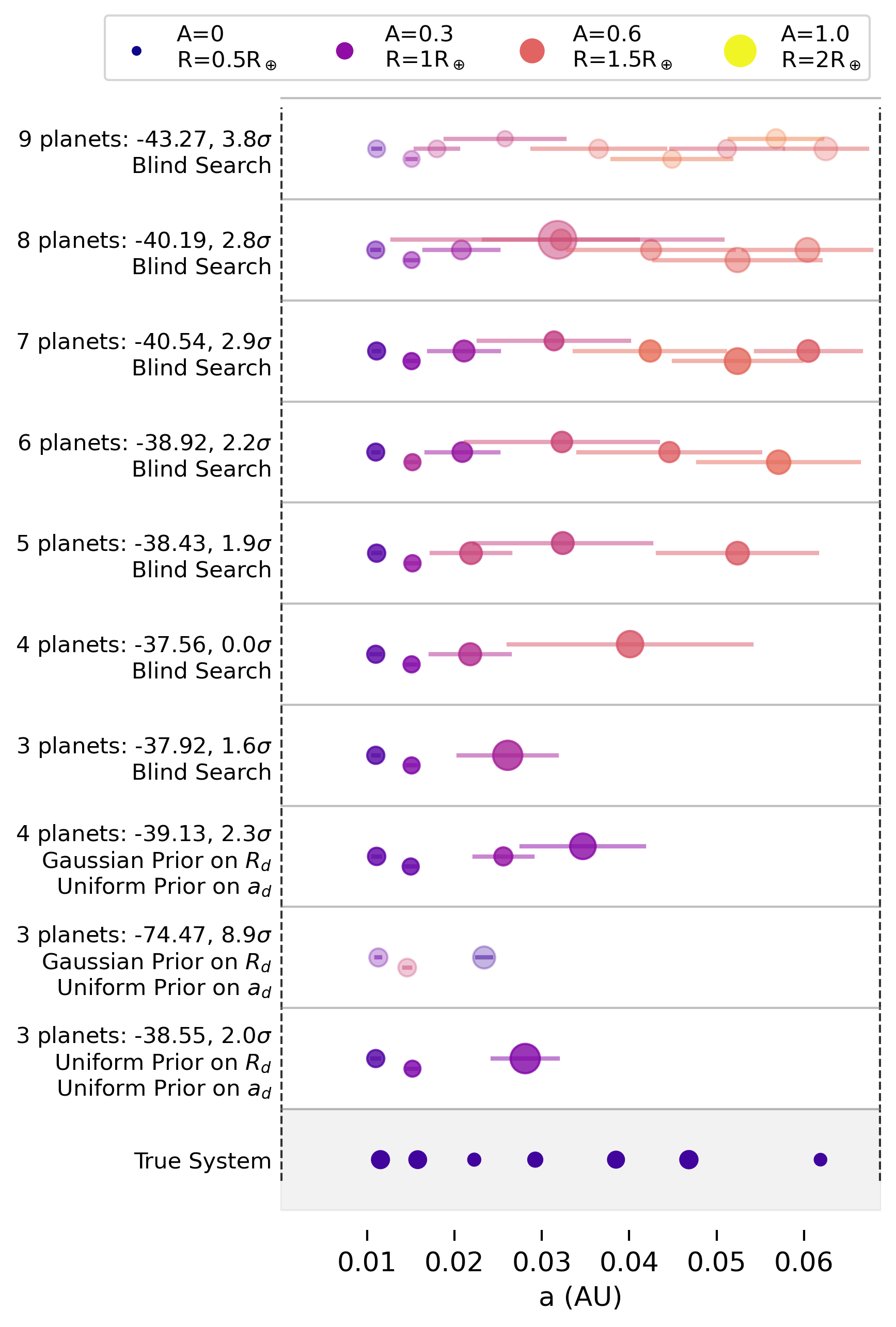}
    \caption{Same as \autoref{fig:2015mods} but using the information gained from \citet{Gillon2016}. Despite the fact that the eight-planet case is within 3$\sigma$, the planets ``e" and ``f" are trying to compensate for each other and thus fundamentally show little difference from the seven-planet model.}
    \label{fig:2016mods}
\end{figure}

\citet{Gillon2016} reported the discovery of b, c, and a candidate planet ``d" in the TRAPPIST-1 system. We take the best-fit parameters and their errors for planets b and c and use Gaussian priors on their radii and semi-major axes. We use uniform priors on their albedos as before ($U$(0,1)). \citet{Gillon2016} gave a best fit radius based on their best fit period of 18.202\,days but gave the full range of possible semi-major axes. Given the uncertainty on candidate planet ``d," we opt to model several cases. First, we model the case where we take a uniform prior on the radius ($U$(0.5-4)\,$R_\oplus$) and a uniform prior on the semi-major axis ($U$(0.022,0.146)\,AU). Note that this semi-major axis range is larger than our blind search range of $U$(0.000212,0.068709)\,AU. Second, we model the case where we use the \citet{Gillon2016} Gaussian prior on the radius and uniform prior on the semi-major axis. Third, we model the blind-search case where we use our semi-major axis prior and a uniform prior on the radius. Finally, motivated by our 2015 prediscovery analyses, we also test adding planets up to a system size of nine and examine the evidence of all of the models. The best-fit model of each scenario is shown in \autoref{fig:2016mods}.

The results from these analyses place our PIE insights in contention with the \citet{Gillon2016} best guess for candidate planet ``d." Their best guess for candidate planet ``d" placed it at a period more consistent with real planet h, while our result is more consistent with the actual planet d (when we use their Gaussian prior on the radius). It is interesting to note, in retrospect, how the suggested orbital periods for candidate ``d" in \citet{Gillon2016} are similar to, but not entirely consistent with, the eventual orbital periods of the real planets d, e, f, g, and h. Ideally, we would have liked to test each of their best-fit cases with each of their subsequent best-fit radii and semi-major axes, but we are limited by how they reported potential candidate ``d" parameters. 

The best-fit model based on the evidence is four planets in the system when we ignore the tentative information on candidate ``d"; however, the Bayes factors that we compute for each of the other models in comparison to this four-planet case continue to allow for two to seven planets in the system. \citet{Gillon2016} use 2550\,K for the stellar temperature and 0.117\,$R_\odot$ for the stellar radius. We maintain that the star is 0.1192\,$R_\odot$ and 2566\,K. We use the new stellar mass from \citet{Gillon2016} (0.080\,$M_\odot$) and assume circular orbits to determine the best-fit periods for planets ``d" and ``e." We would report periods of 4.1$^{+1.3}_{-1.2}$\,d and 10.2$^{+5.8}_{-4.8}$\,d (the \citealp{Agol2021} values are 4.049219 and 6.101013, respectively). In all favored cases, the true period of planet d is within 2$\sigma$ of our prediction. In all favored cases where a planet ``e" exists, the true period of planet e is within 1$\sigma$. Thus a PIE analysis conducted in 2016 would have ruled out the \citet{Gillon2016} suggested period for candidate planet ``d" and provided constraints on future planet searches.

\subsection{Post-Spitzer} \label{sec:spitzer}
\begin{figure}
    \centering
    \includegraphics[width=\linewidth]{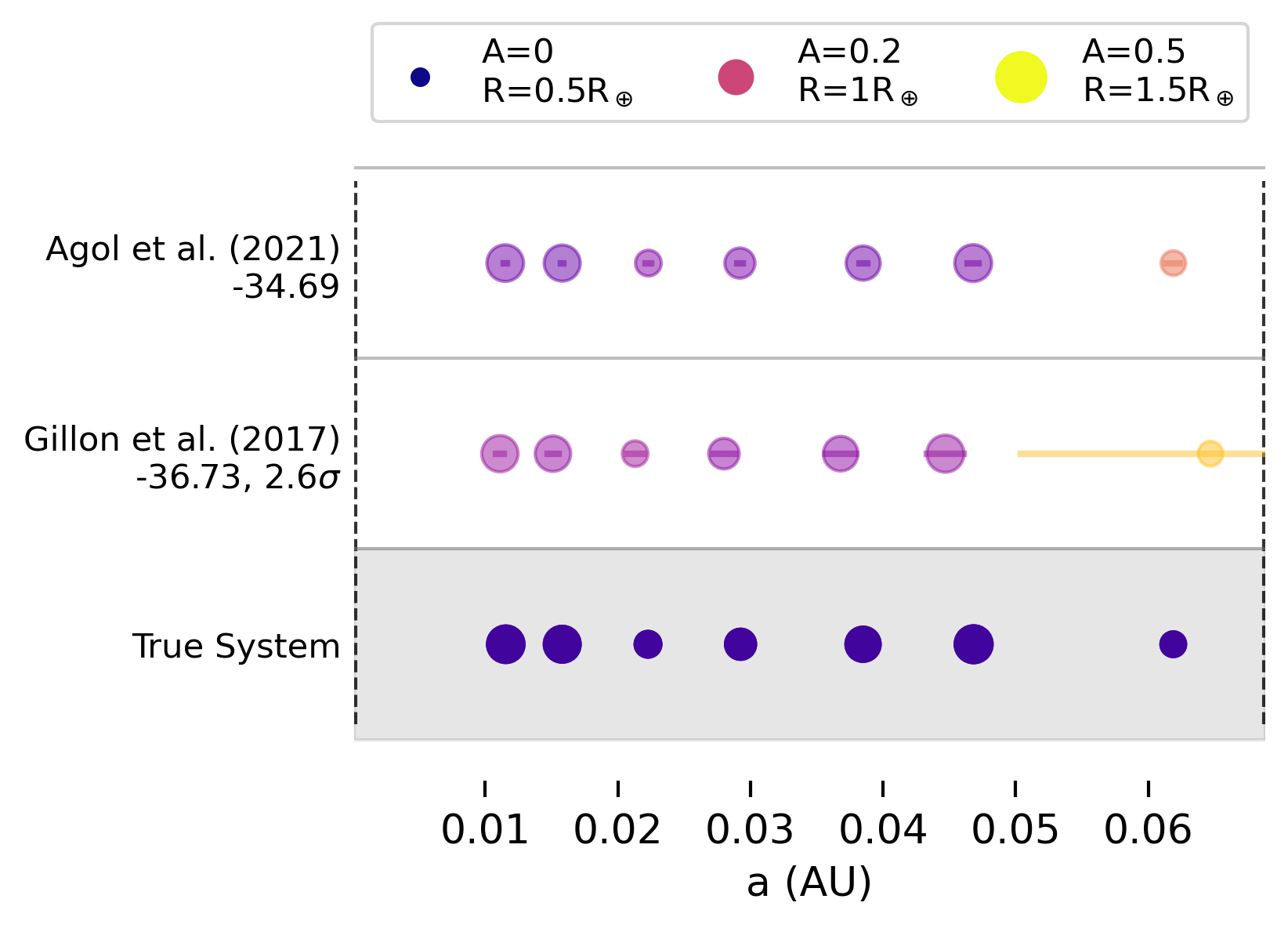}
    \caption{Best-fit system in each constraint scenario compared to the true TRAPPIST-1 system. The symbols are sized by the planet radius and colored by the albedo. Shaded bars indicate the error on the planet's semi-major axis. Each modeling scenario is labeled with the source of the priors, the computed evidence, and for the \citet{Gillon2017} scenario we show the significance compared to the \citet{Agol2021} best-fit model. The posteriors do not improve over the priors in either case except for the upper limit on planet h's semi-major axis (and subsequent orbital period), which has a more precise upper limit.}
    \label{fig:seven}
\end{figure}

\citet{Gillon2017} reported the presence of seven planets around TRAPPIST-1 following extensive photometric follow-up that featured an impressive 20 days of continuous monitoring with Spitzer. The parameters on planet h are relatively uncertain, and the suggested semi-major axis upper limit lies outside of our blind-search distance, and thus, we expect PIE to struggle to constrain planet h's parameters. Then, \citet{Luger2017} used K2 observations to refine planet h's parameters, and finally we arrive at the present best known parameters of \citet{Agol2021}, a photodynamical analysis. We test a 7-planet retrieval using the \citet{Gillon2017} constraints and the \citet{Agol2021} constraints (which were used to create the ``data" and thus represent the truth) and show the best-fit results in \autoref{fig:seven}. We see that PIE struggles to constrain planet h's semi-major axis using the \citet{Gillon2017} priors, but our upper limit on the orbital period is more tightly constrained (21$^{+7}_{-6}$\,days vs 20$^{+15}_{-6}$\,days). All other planets have consistent parameters with no remarkable differences in precision. With the tight constraints provided by \citet{Agol2021}, PIE produces no improved constraints on the parameters of $A$, $a$, or $R$.

\section{Discussion} \label{sec:discuss}
With a MIRECLE-like telescope, it is clear in the case of TRAPPIST-1 that the star is not alone but has excess flux indicative of planets as demonstrated by \autoref{fig:zero}, where the residuals using known values in 2015 as priors exceed the noise floor compared to when fitting a planet-free star. Our additional analysis resulted in a prediction that the star, then 2MASS J23062928-050228, was planet hosting, with two to seven planets. The ``2015" case demonstrates that a MIRECLE-like mission with precise observations could use the PIE technique to reveal non-transiting TRAPPIST-1-like systems. The PIE technique could be used to identify planet-hosting candidates and constrain the potential number of exoplanets in the system. Luckily, the TRAPPIST-1 planets do transit and allow us to gain additional insights as to the performance of the PIE technique in such compact multiplanet scenarios.

\begin{figure*}
    \centering
    \includegraphics[width=\linewidth]{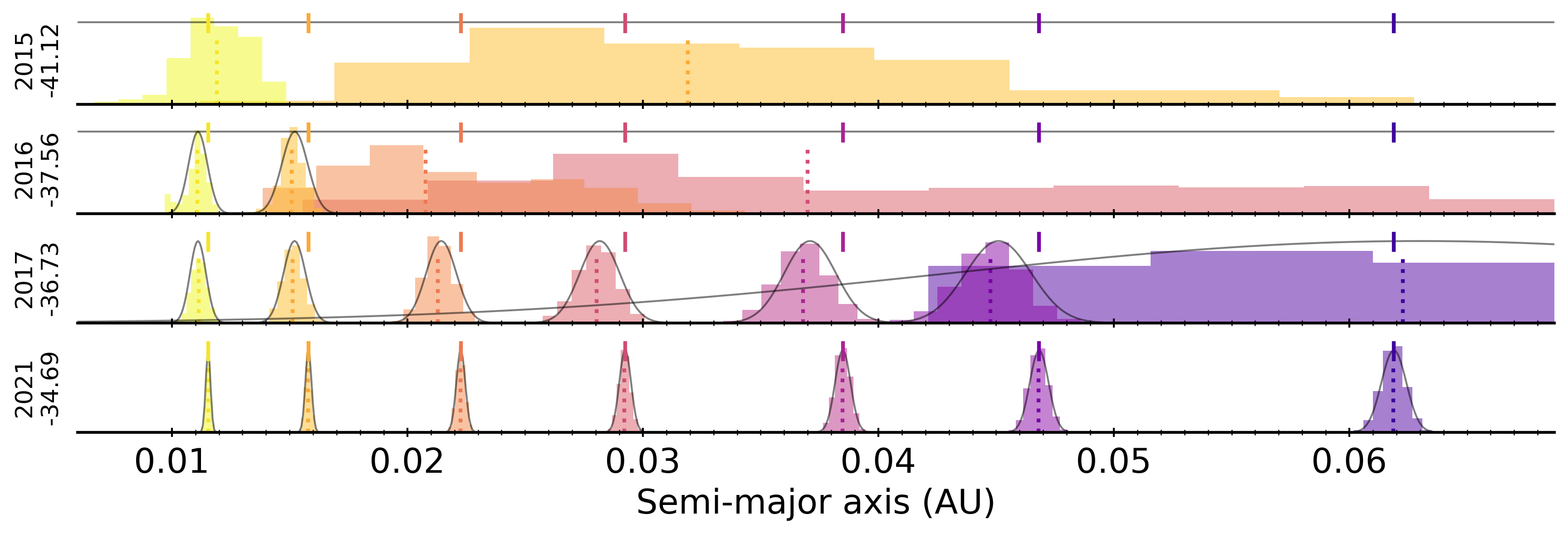}
    \caption{The one-dimensional priors and posteriors for the retrieved semi-major axes for the best-fit models throughout our historical analysis. The posteriors for each planet are given in a different color with the prior denoted by a gray envelope. Solid tick marks show the true \citet{Agol2021} values, and the dotted lines mark the mean retrieved values. In ``2015" and ``2016," we used uniform priors that extend beyond the plotted region for any unconfirmed planets; thus, only the outline (a flat gray line) appears in the figure. Alongside the year labels, we show the evidence.}
    \label{fig:smahist}
\end{figure*}

\begin{figure*}
    \centering
    \includegraphics[width=\linewidth]{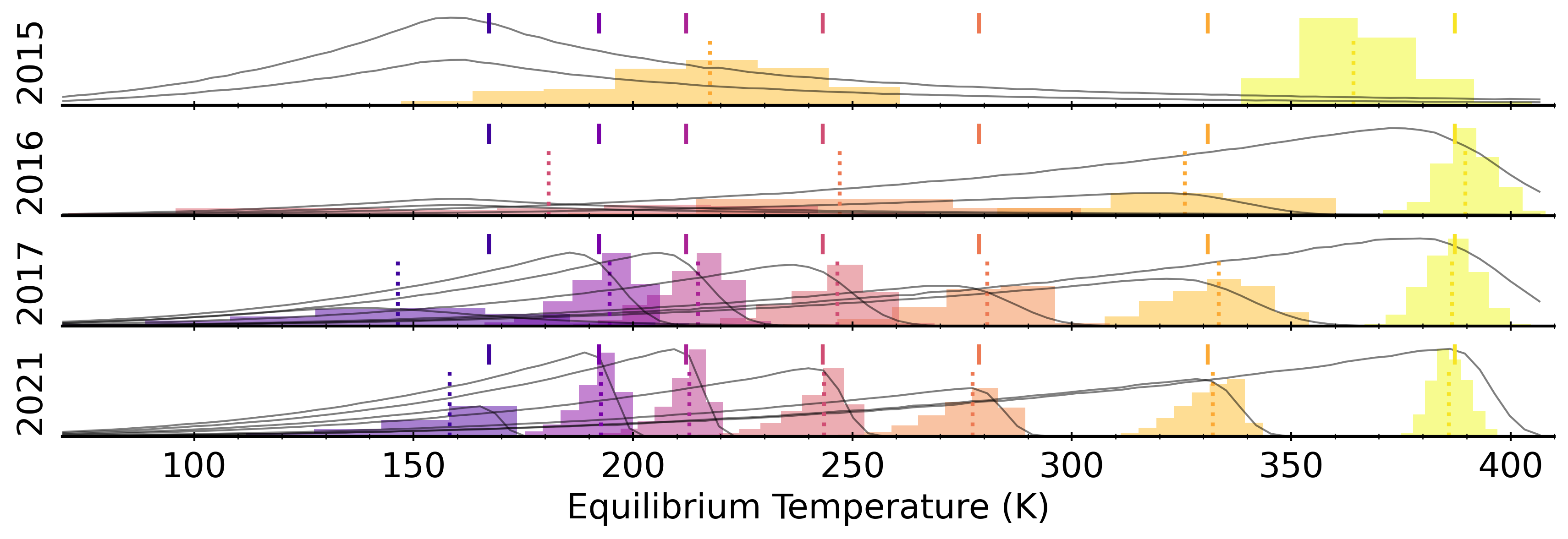}
    \caption{Same as \autoref{fig:smahist} but for the equilibrium temperatures of the included planets for the best-fit models throughout our historical analysis. Unlike the prior reliant semi-major axes, the PIE technique can place informative constraints on the equilibrium temperatures of the planets.}
    \label{fig:eqthist}
\end{figure*}

Because we have computed the evidence for each of our best-fit models, we can readily compare how our inferences for the system evolve with our knowledge of the TRAPPIST-1 system. We examine the 1D posteriors for the retrieved semi-major axis of our best-fit models through the TRAPPIST-1 timeline in \autoref{fig:smahist}. The final best fit with \citet{Agol2021} priors had an evidence of --34.69, a 4.7$\sigma$ improvement from our initial blind search for seven planets (--44.0). Despite the fact that the \citet{Agol2021} assumed system parameters were included in the blind search region implemented, it was not the best-fit solution. This is because the posteriors of the ``2015" seven-planet best fit are multimodal and several parameter combinations achieve similar fits. Thus, we see here that in a compact, multiplanet case like TRAPPIST-1, the PIE technique struggles to constrain the semi-major axes of the planets without strong nonoverlapping priors.  This can be seen in \cref{fig:2015mods,fig:2016mods,fig:smahist,fig:eqthist}, where the semi-major axis uncertainties often extend to each planet's neighbor.

The posteriors often show a correlation between semi-major axes for neighboring planets. This effect is driven by the constraints we imposed on the semi-major axes. Note that to avoid identity confusion, we specified that the semi-major axes must occur in order of smallest to largest. In our initial conception of the experimental design, we had chosen to fit for the equilibrium temperatures, $T_i$, and use them for identity ordering. This choice, however, leads to a faux prior boundary on the temperatures. Since the equilibrium temperatures of the TRAPPIST-1 planets are within 30\,K of each other, if the temperature for planet b, for example, starts too cold, it artificially drives the other planets to be too cold as well and thus, forces the model to increase their radii to compensate. However, this effect is still seen when, for example, the prior for planet g is well constrained and the prior for planet h is not, as seen in ``2017." Here, the posterior for the semi-major axis of planet h is Gaussian with a cutoff on the short end due to this constraint by planet g's semi-major axis. The advantage of using $A$ and $a$ instead is that together these two parameters can compensate to match the right effective temperature. Additionally, this design choice is more physical in that no planets can be in the same position, but, due to surface and atmospheric properties, they can be the same temperature or not correlated with distance from the host star.

Despite this choice, we still see correlations between the semi-major axes of subsequent planets. We can evaluate this more by computing the Pearson's and Spearman's correlation coefficients, where values greater than 0.5 indicate a strong correlation. Using our four-planet model from ``2015," we see a correlation (Spearman's = 0.573) between $a_c$ and $a_d$ (planet's N and N-1). The five-planet model shows a correlation between $a_d$ and $a_e$ and $a_e$ and $a_f$ (N and N-1, N-1 and N-2). At six-planets, the correlation stretches from N to N-5 and finally includes all planets once we hit eight-planets in the system. In ``2016," correlations reach the 0.5 threshold at six-planets and include N to N-3.

\begin{figure*}
    \centering
    \includegraphics[width=\linewidth]{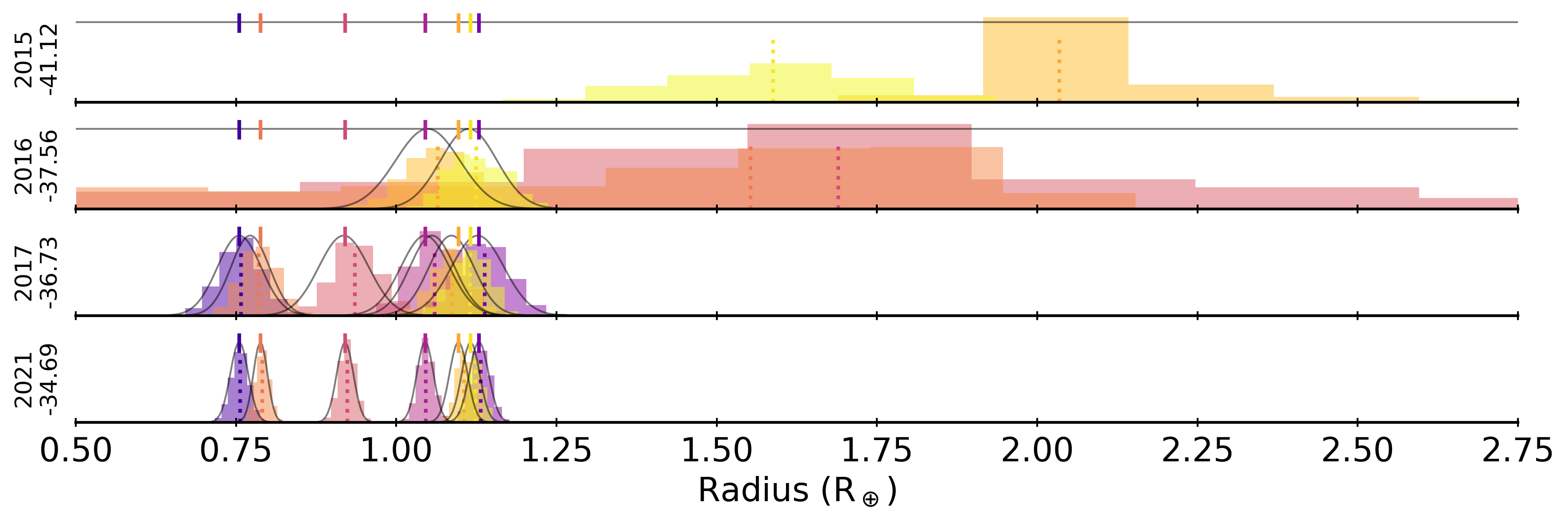}
    \caption{Same as \autoref{fig:smahist} but for the radii of the included planets for the best-fit models throughout our historical analysis. Unlike the prior reliant semi-major axes, radius can shift from the priors to compensate for $A$ and $a$.}
    \label{fig:radhist}
\end{figure*}

In \autoref{fig:eqthist}, we examine the equilibrium temperatures of the planets in our best-fit models through ``time" (2015 -- 2021). Despite the strong dependence the PIE technique has on the semi-major axis priors, having albedo be a free parameter allows us to estimate the true equilibrium temperatures of the planets. Thus in 2017 and 2021, the PIE technique is capable of retrieving the planets' equilibrium temperatures. This is additional information gained even when the system is well characterized in terms of known planets and their rough parameters. While the albedo and the semi-major axis for each planet ultimately control where the peak of the contributed flux occurs in wavelength space, the radius of the planet has much more control over how much flux is being added to the system's flux. In \autoref{fig:radhist}, we examine the radii of the planets in our best-fit models through ``time" (2015 -- 2021). The radius of the planet is thus less prior driven than the semi-major axis and ends up being larger to compensate for any incorrect combinations of $A$ and $a$.

\subsection{Variation on Model Assumptions} \label{sec:jwst}
\begin{figure*}
    \centering
    \includegraphics[width=\linewidth]{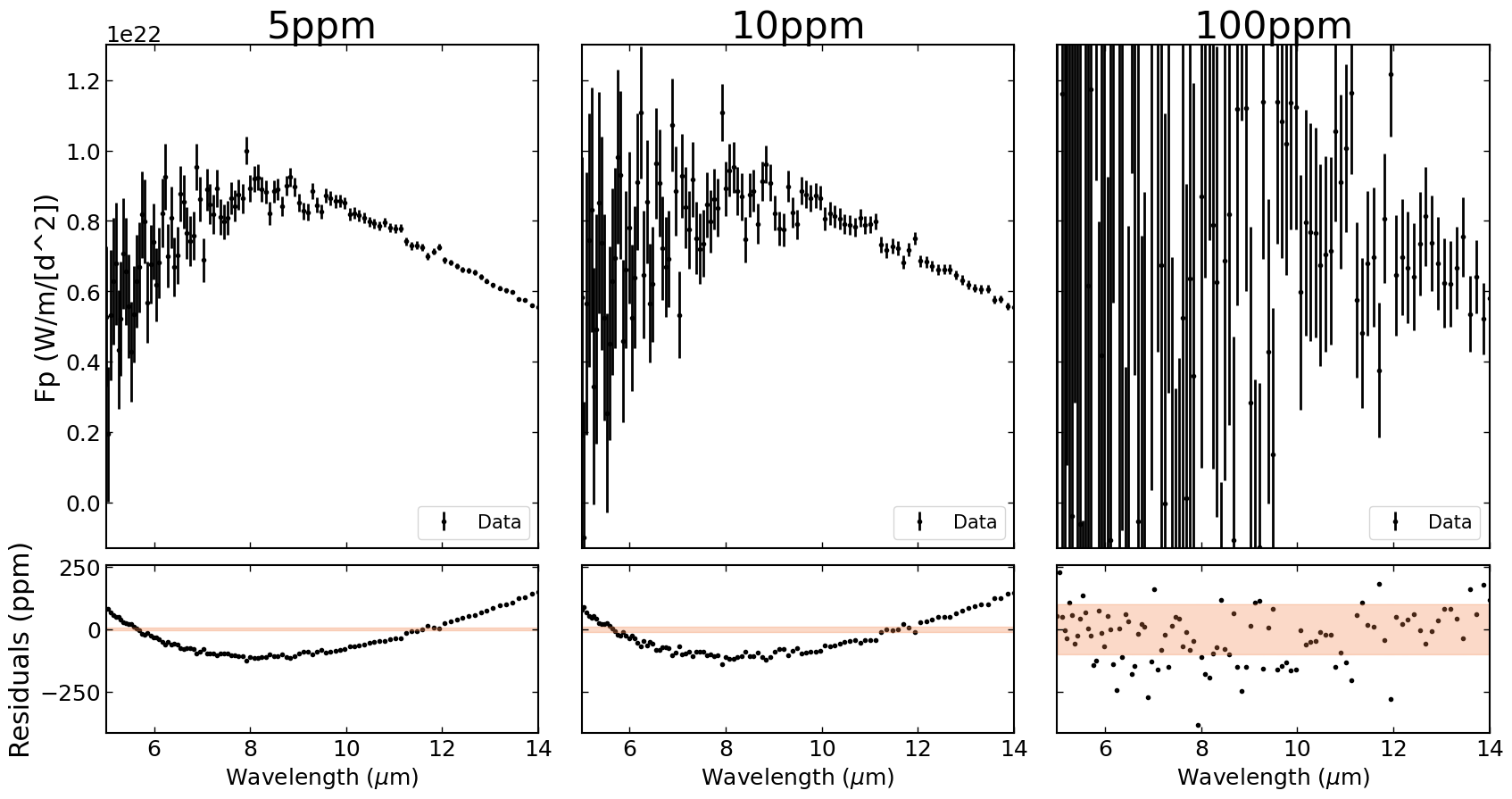}
    \caption{The planetary fluxes of the true system in the noisy JWST/MIRI-LRS scenarios (left: 5\,ppm, middle: 10\,ppm, right: 100\,ppm). Top: the flux from the planets. Bottom: the residuals when fitting a star-only model compared to the corresponding noise floor (orange shaded region) for each scenario.}
    \label{fig:jwst-star}
\end{figure*}

Our MIRECLE exploration is in many ways highly idealized, with its approximately noiseless construction and a sufficiently broad wavelength coverage so as to capture the peak of all the planets' blackbody functions. A study by \citet{Lustig-Yaeger2021} previously showed that JWST was effective for the PIE technique using realistic planetary atmospheres in single hot-Jupiter planet systems, while we have constrained ourselves to blackbodies. Another study tested white-dwarf host stars \citep{Limbach2022} instead of M-dwarf hosts, and we have neglected stellar activity and used a blackbody. We also explored more realistic scenarios (5\,ppm, 10\,ppm and 100\,ppm noise) with JWST/MIRI-LRS style observations, with a wavelength range of 5--14\,$\mu$m and a spectral range, $R=100$ with the TRAPPIST-1 architecture, maintaining our blackbody assumptions. The limited wavelength range yields a truncated prior on the semi-major axis of 0.005--0.041\,AU, and thus, we predict that we would only be sensitive to the five innermost planets in the TRAPPIST-1 system. For these scenarios, rather than imposing an artificial noise floor of 5\,ppm, we add noise to the spectrum. For easy comparison across all noise levels, we first computed a 1\,ppm noise realization using a normal distribution relative to the stellar flux and then scaled that so that all the noise realizations would be the same and comparable. For example, we show the residuals for each of the star-only initial searches for the JWST case in \autoref{fig:jwst-star}. In the pseudo-noiseless case with MIRECLE, we accurately retrieved the temperature and radius of the star in every exploration. With noise, the stellar parameters now drift toward a cooler and larger star as expected.

\begin{figure*}
    \centering
    \includegraphics[width=\linewidth]{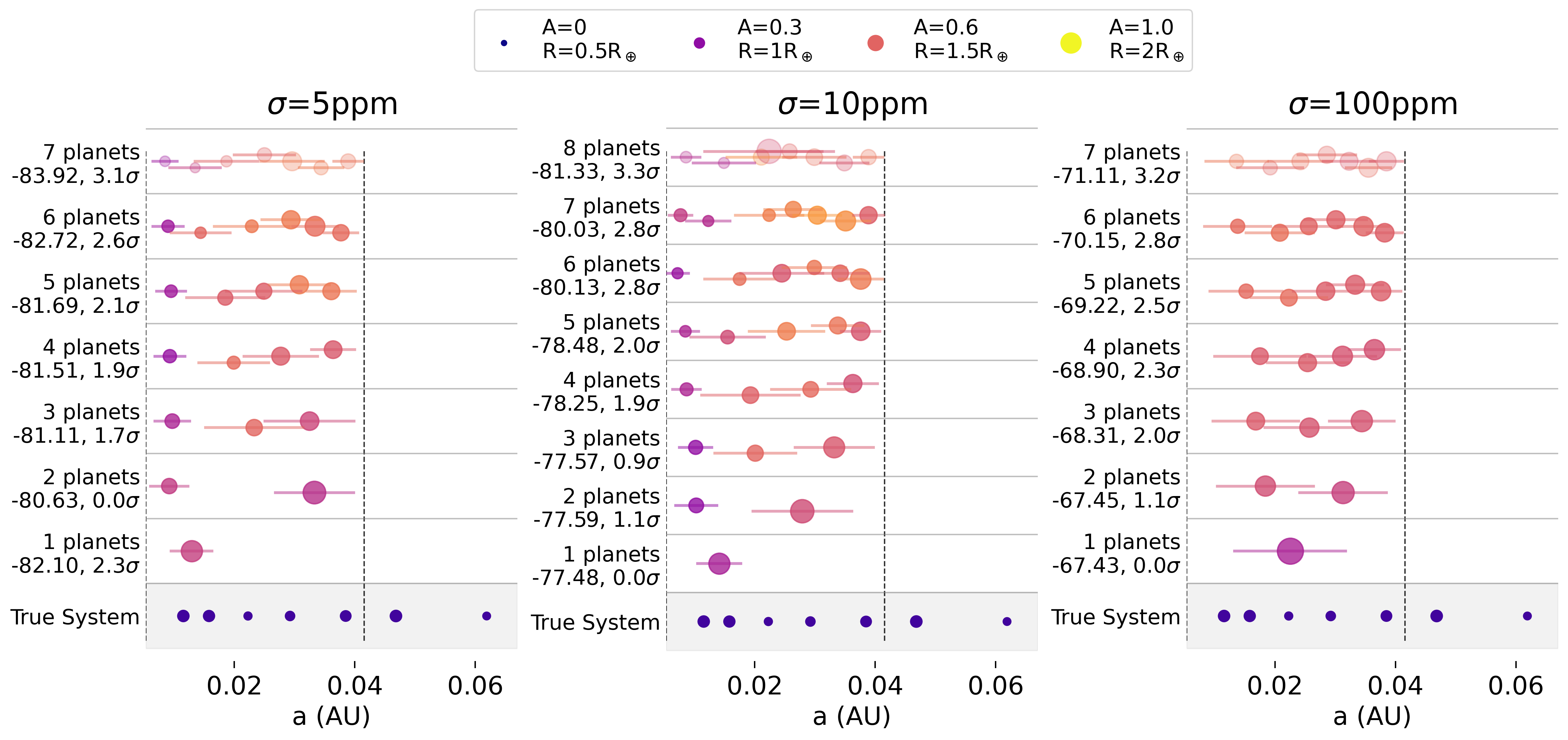}
    \caption{Same as \autoref{fig:2015mods} but with 5\,ppm noise (left), 10\,ppm noise (middle), and 100\,ppm noise (right) with JWST/MIRI-LRS observations. Since we do not allow for semi-major axes beyond our 0.041 AU upper limit, planets g and h are not represented in any of our best-fit solutions.}
    \label{fig:2015noise}
\end{figure*}

A ``2015" blind search from planets in the system reveals the effects of both the wavelength cutoff and the noise, and we show our best fits in \autoref{fig:2015noise}. With MIRECLE in ``2015," we had placed a lower limit on the number of planets at two, and that is no longer the case for the JWST scenarios, but it is the preferred solution in the 5\,ppm highest precision case. Larger noise realizations prefer the one-planet system scenario. In all of our JWST scenarios, the evidence prefers at least one planet and rules out the planet-free scenario at $>$5$\sigma$ confidence. In the 5 and 100\,ppm scenarios we would allow up to six planets in the system, but 10\,ppm would allow seven planets in the system. Note that we do not evaluate the stability of the system configurations.

\begin{figure*}
    \centering
    \includegraphics[width=0.32\linewidth]{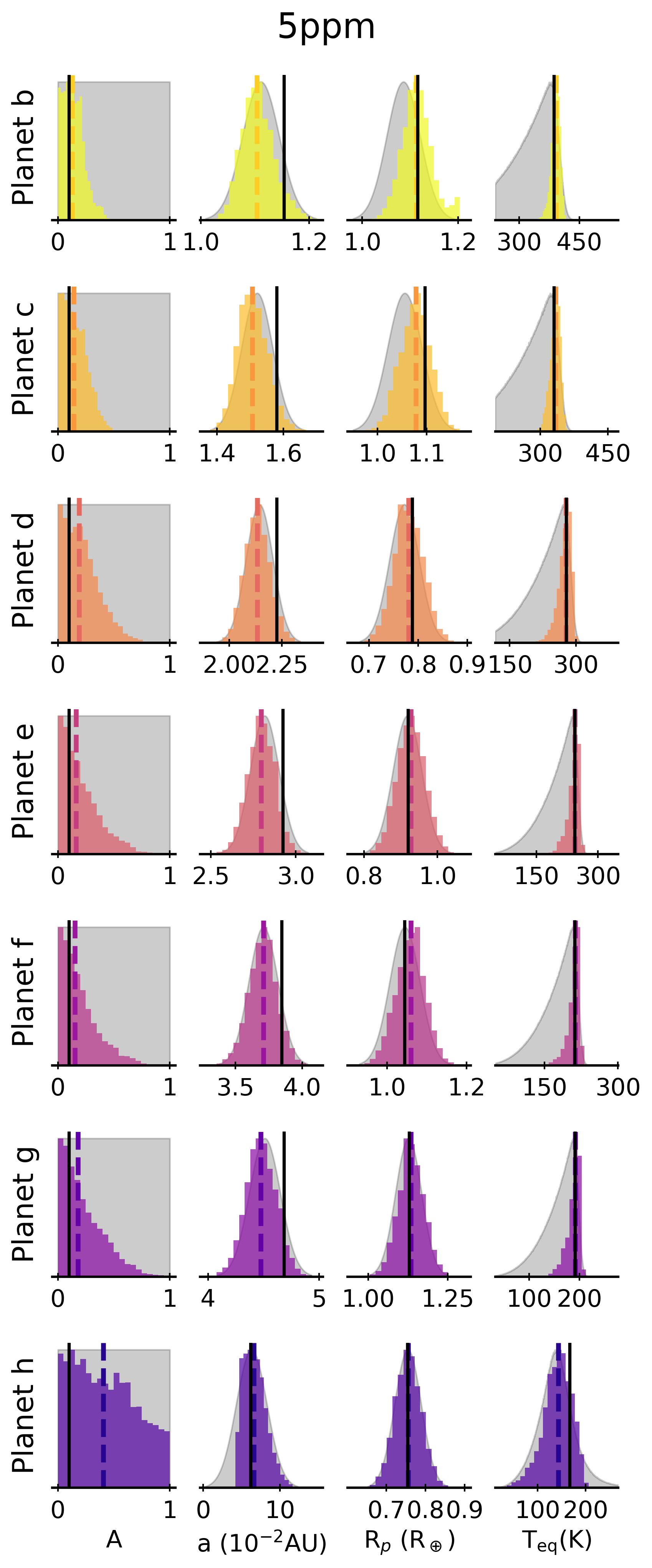}
    \includegraphics[width=0.32\linewidth]{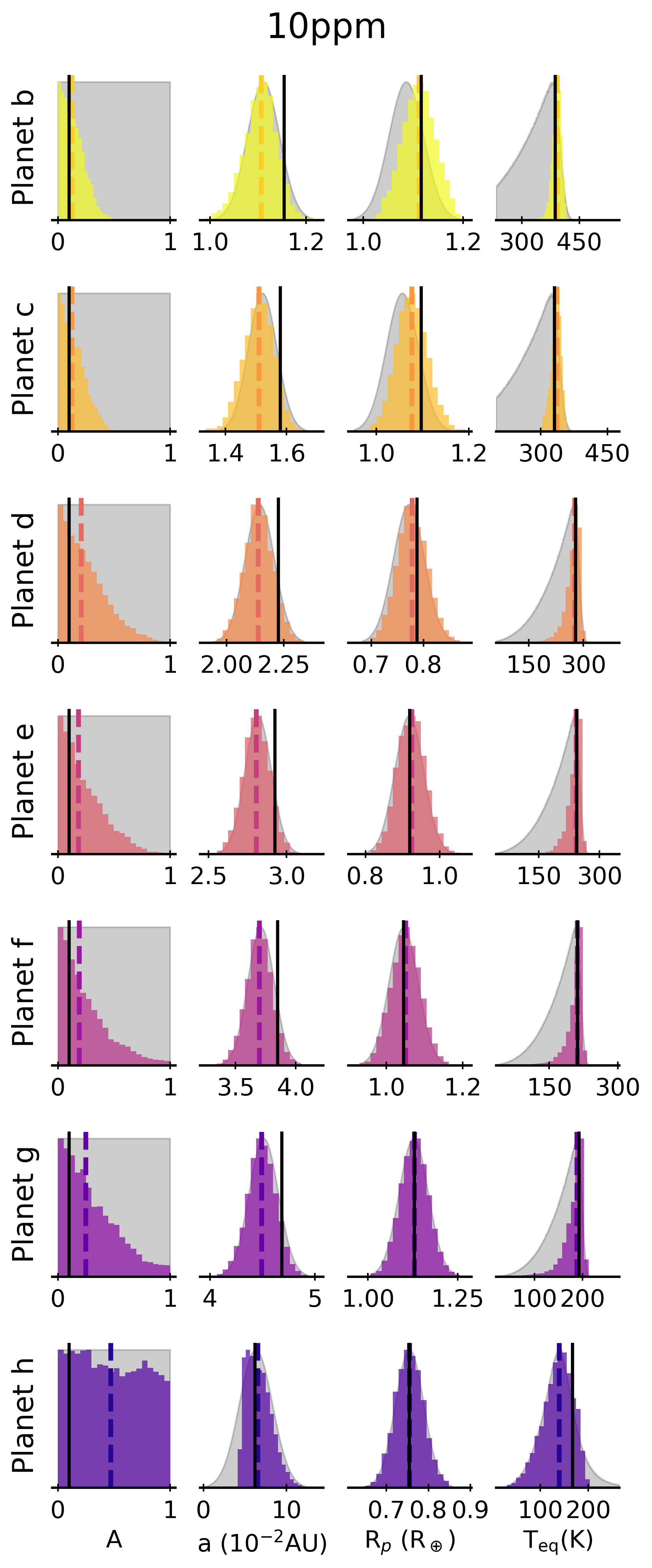}
    \includegraphics[width=0.32\linewidth]{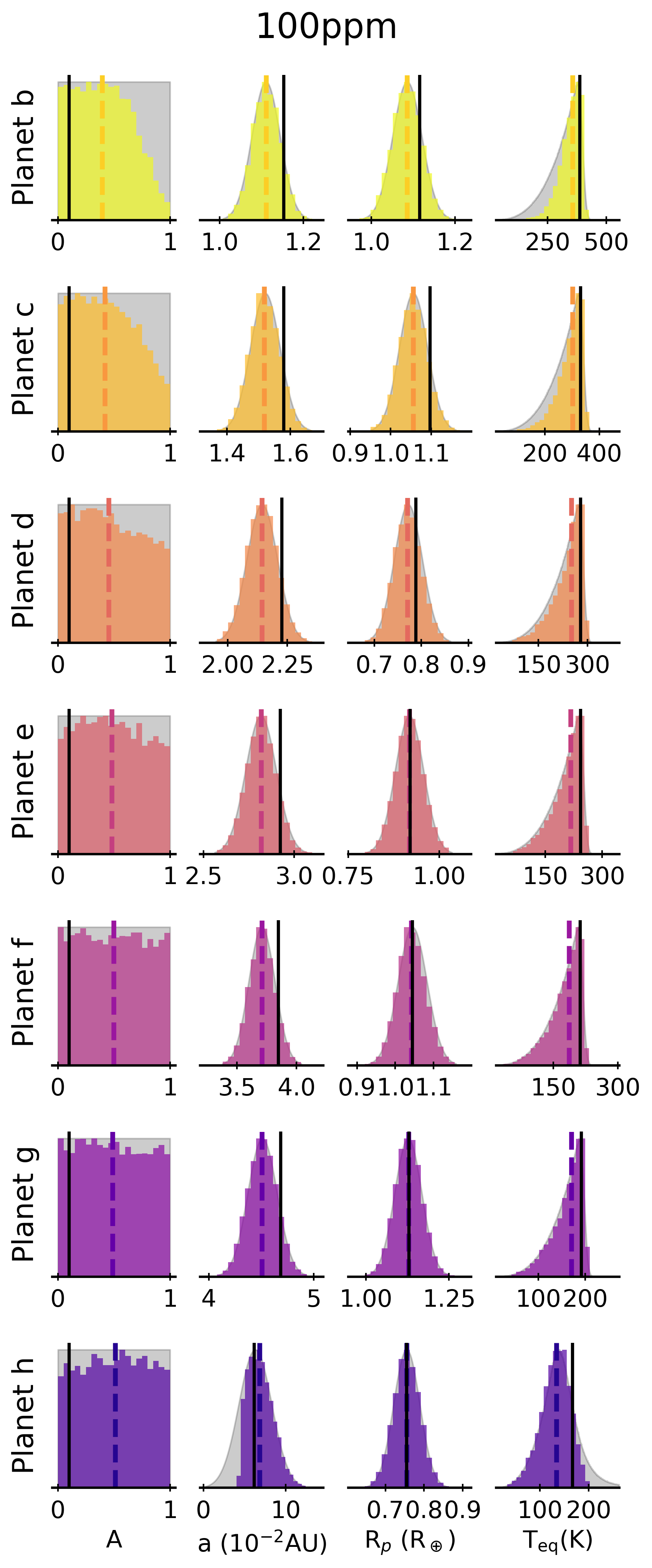}
    \caption{The 1D posteriors and priors \citep[from][]{Gillon2017} for the three noise renditions with simulated JWST data (left: 5\,ppm, middle: 10\,ppm, right: 100\,ppm). The posteriors for each planet (top row: planet b) are given in different colors. Each column is a different fit parameter with the post-processed equilibrium temperature shown on the rightmost column of each noise rendition. The priors are shown in gray with the ``true" value \citep[from][]{Agol2021} shown as a black solid line and the median of the posterior shown as a colored dashed line.}
    \label{fig:jwstpriors}
\end{figure*}

Despite not being sensitive to the peaks of planet g and h's blackbody flux, knowing that they exist allows us to see if JWST/MIRI-LRS data can infer their temperatures. Using the \citet{Gillon2017} priors, we fit for all seven planets with the three noise renditions and compared our results to the priors to see what parameters are constrained and what information is gained by PIE JWST observations. We show this in \autoref{fig:jwstpriors}. Comparing across planets, there is little difference in the precision of the constraint on the temperature between planets e, f, and g, but planet h is no more constrained than by the initial prior with the exception of the truncation of the posterior at roughly 200\,K. This is because of the semi-major axis prior, where we require that the semi-major axes of the planets always occur in order from smallest to largest. Thus, even though the prior is broad, semi-major axes smaller than that of planet g are not allowed for planet h. This is true at all noise levels.

In general we see that at the largest noise level, the posteriors are strongly prior dependent for both the semi-major axis and the radius. As the precision improves, the radius becomes less prior dependent and PIE is able to find the more appropriate planet radius. Higher precision also allows for better limits on the planet's albedo, where in the highest-precision rendition we begin to place functional upper limits on the albedos of several planets.

\section{Conclusions} \label{sec:conc}
We have presented a demonstration on the effectiveness of the PIE technique \citep{Stevenson2020} on the tightly packed multiplanet system of TRAPPIST-1. We simulated two telescope observation scenarios, JWST and MIRECLE, and tracked how additional knowledge assisted PIE in constraining planetary parameters. We find that even in the case where no planet information is known, a case equivalent to a non-transiting system, the PIE technique is able to determine if planets are present. The blind search for planets is sensitive to the wavelength range of observation. However, we find that in this multiplanet scenario, PIE struggles to constrain the semi-major axes of all the planets without strong priors. Despite this, PIE still can be used to constrain the equilibrium temperatures of known planets in the system even if their blackbody flux peaks outside of the observed wavelength range.

The MIRECLE and JWST observation cases presented here are idealized. In reality, we would require hundreds of hours of JWST time, and the data would need to bin down cleanly (i.e., no astrophysical or systematic red noise) to obtain such precise data with TRAPPIST-1. However, the solar neighborhood is full of M dwarfs that are much brighter and quieter than TRAPPIST-1, and thus more amenable to observations. Additionally, we did not consider the distance to the system in our fits; we take that as a known quantity, and observations with any facility would need to take into account such flux affecting quantities. Fortunately, nearby stars benefit from having comparatively precise parallax constraints \citep[from missions like Gaia;][]{Gaia}, which will benefit the determination of their distances. We also constrained our analysis to blackbody representations of all bodies in the system. In reality, the atmospheres (or lack thereof) of all bodies in the system and their variability would affect our ability to constrain the planets' temperatures. Future work will look to explore the addition of atmospheres and a realistic stellar spectrum on the effects of multiplanet PIE.

\section*{acknowledgments}
L.C.M would like to acknowledge K-pop, ramyeon, bubble tea, and Keven Cetaceanson for the physical and emotional support necessary to complete this work.

This research has made use of the NASA Exoplanet Archive, which is operated by the California Institute of Technology, under contract with the National Aeronautics and Space Administration under the Exoplanet Exploration Program.

This material is based upon work performed as part of the CHAMPs (Consortium on Habitability and Atmospheres of M-dwarf Planets) team, supported by the National Aeronautics and Space Administration (NASA) under grant No. 80NSSC21K0905 issued through the Interdisciplinary Consortia for Astrobiology Research (ICAR) program.

\software{dynesty \citep{dynesty}, astropy \citep{Astropy}, matplotlib \citep{Matplotlib}, pandas \citep{Pandas}, numpy \citep{NumPy}, PySynphot \citep{PySynphot}, SciPy \citep{SciPy}, corner \citep{Foreman-Mackey2016}}

\bibliography{main}{}
\bibliographystyle{aasjournal}

\end{document}